\documentclass[fleqn,12pt,twoside]{elsart}

\usepackage{graphicx}
\usepackage{amssymb,amsmath}

\usepackage{citesort}

\newcommand{\be}{\begin{eqnarray}}
\newcommand{\ee}{\end{eqnarray}}

\newcommand{\pme}[2]{\ensuremath{\,\pm\,{}^{#1}_{#2}}}

\newcommand{\beq}{\begin{equation}}
\newcommand{\eeq}{\end{equation}}
\newcommand{\bea}{\begin{eqnarray}}
\newcommand{\beas}{\begin{eqnarray*}}
\newcommand{\beau}[1]{\begin{equation} \label{#1} \begin{array}{rcl}}
\newcommand{\eea}{\end{eqnarray}}
\newcommand{\eeas}{\end{eqnarray*}}
\newcommand{\eeau}{\end{array} \end{equation}}
\newcommand{\bay}{\begin{array}}
\newcommand{\eay}{\end{array}}
\newcommand{\bals}{\begin{align*}}
\newcommand{\eals}{\end{align*}}

\begin{document}

\begin{frontmatter}



  \title{ Erratum: Atomic Mass Dependence of Hadron Production in
    Deep Inelastic Scattering on  Nuclei}

  \author[Columbia,ISU]{A.~Accardi}
  , \author[Heidelberg]{D.~Gr\"unewald\thanksref{Grunemail}}
  , \author[LNF]{V.~Muccifora} 
  and \author[Heidelberg]{H.J.~Pirner} 


  \address[Columbia]{Columbia Physics Department, 538 West 120th
    Street, New York, NY 10027, U.S.A.}
  \address[ISU]{Iowa State University, Dept. of Physics and
    Astrophysics, Ames, IA 50011, U.S.A.}
  \address[Heidelberg]{Institut f\"ur Theoretische Physik der
    Universit\"at Heidelberg Philosophenweg 19, D-69120 Heidelberg, Germany }
  \address[LNF]{INFN, Laboratori Nazionali di Frascati, I-00044
    Frascati, Italy }
  \thanks[Grunemail]{Corresponding author. E-mail address: daniel@tphys.uni-heidelberg.de}

  
\end{frontmatter}

\setcounter{table}{1}
\setcounter{figure}{7}

\begin{figure}[ht]
\centering
\hspace*{-2.25cm}
\includegraphics[width=18.0cm, height=19.cm]{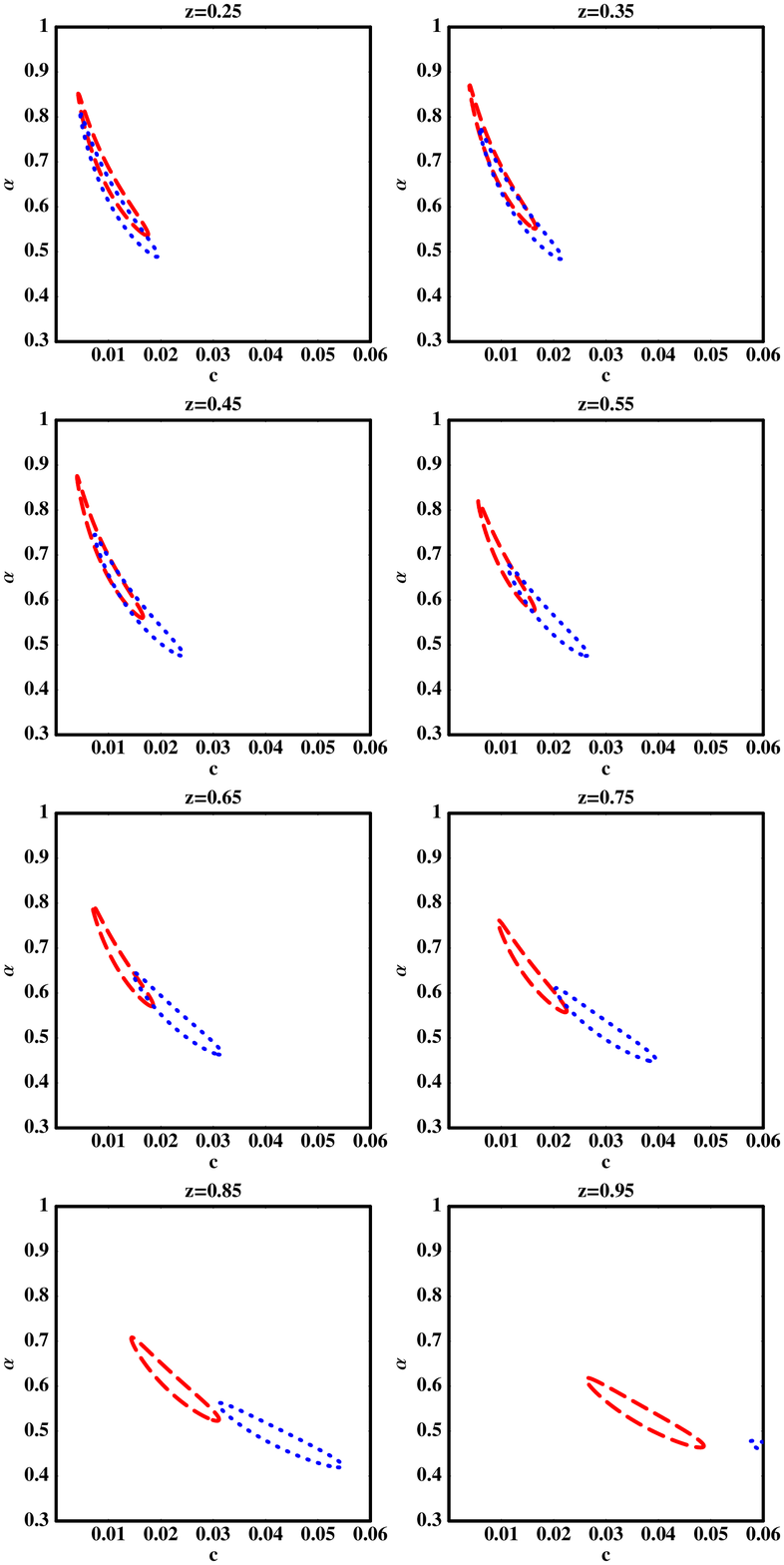}
\vskip-0.5cm
\caption{$\chi^2$ contour plots for the fit $1-R_M=c(\nu,z,h)
  A^\alpha$ on the pure absorption model (dashed) and the full model (dotted) computation for $^{4}$He, ($^{14}$N),
  $^{20}$Ne,  $^{84}$Kr and  $^{131}$Xe nuclei, in fixed $z$-bins. Note that the contour 
  for the full theory computation in the last $z$ bin is outside of
  the plot range, to the right. }
\label{fig:fig7}
\end{figure}

\begin{figure}[tbh]
 \centering
 \includegraphics[width=1.0\linewidth]{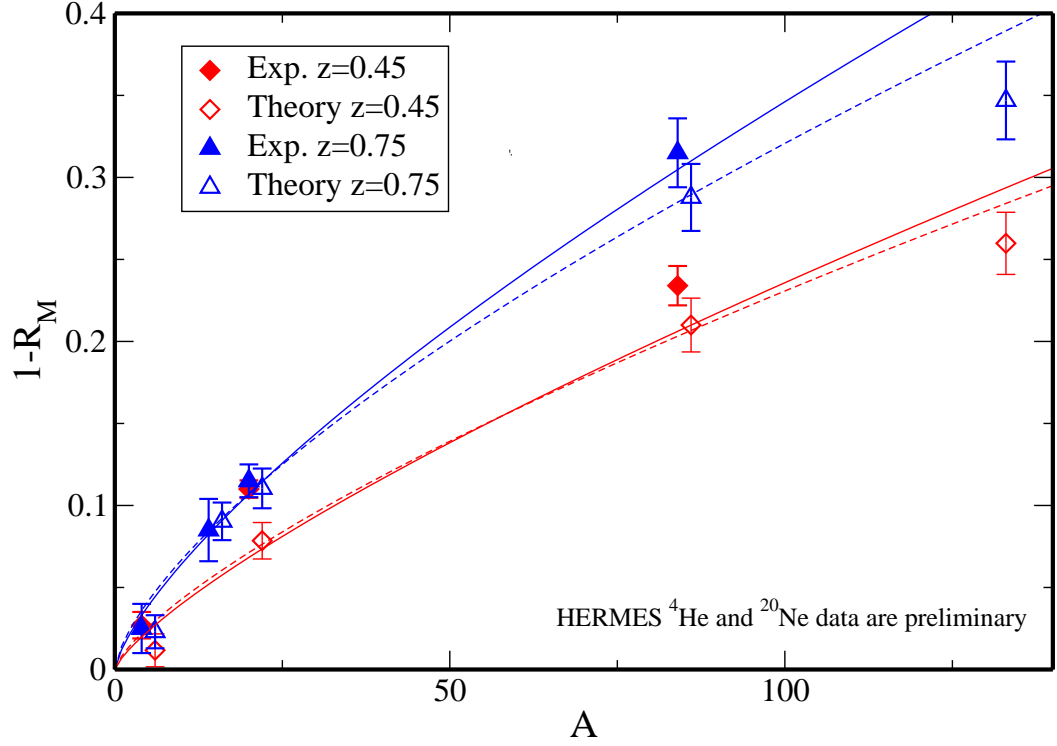}
 \caption{The published HERMES data for $R_M$ for $^{14}$N~\cite{HERM01} 
 and  $^{84}$Kr~\cite{HERM03} and the preliminary data for $R_M$ for 
 $^{4}$He~\cite{HERM04} and $^{20}$Ne~\cite{HERM04} are shown as $1-R_M$ for  
 $z=0.45$ and $z=0.75$ as filled diamonds and filled triangles respectively. 
 The pure absorption model results for $1-R_M$ in the same $z$-bins and
 for the same nuclei plus $^{131}Xe$ are shown by empty symbols.
 The correction affects the dashed lines representing a fit $1-R_M=c A^{\alpha}$ 
 including the Xe nucleus.  
}
 \label{fig:AttenVsA}
\end{figure}

\begin{figure}[h]
\centering
\vskip-1cm
\includegraphics[width=1.0\linewidth]{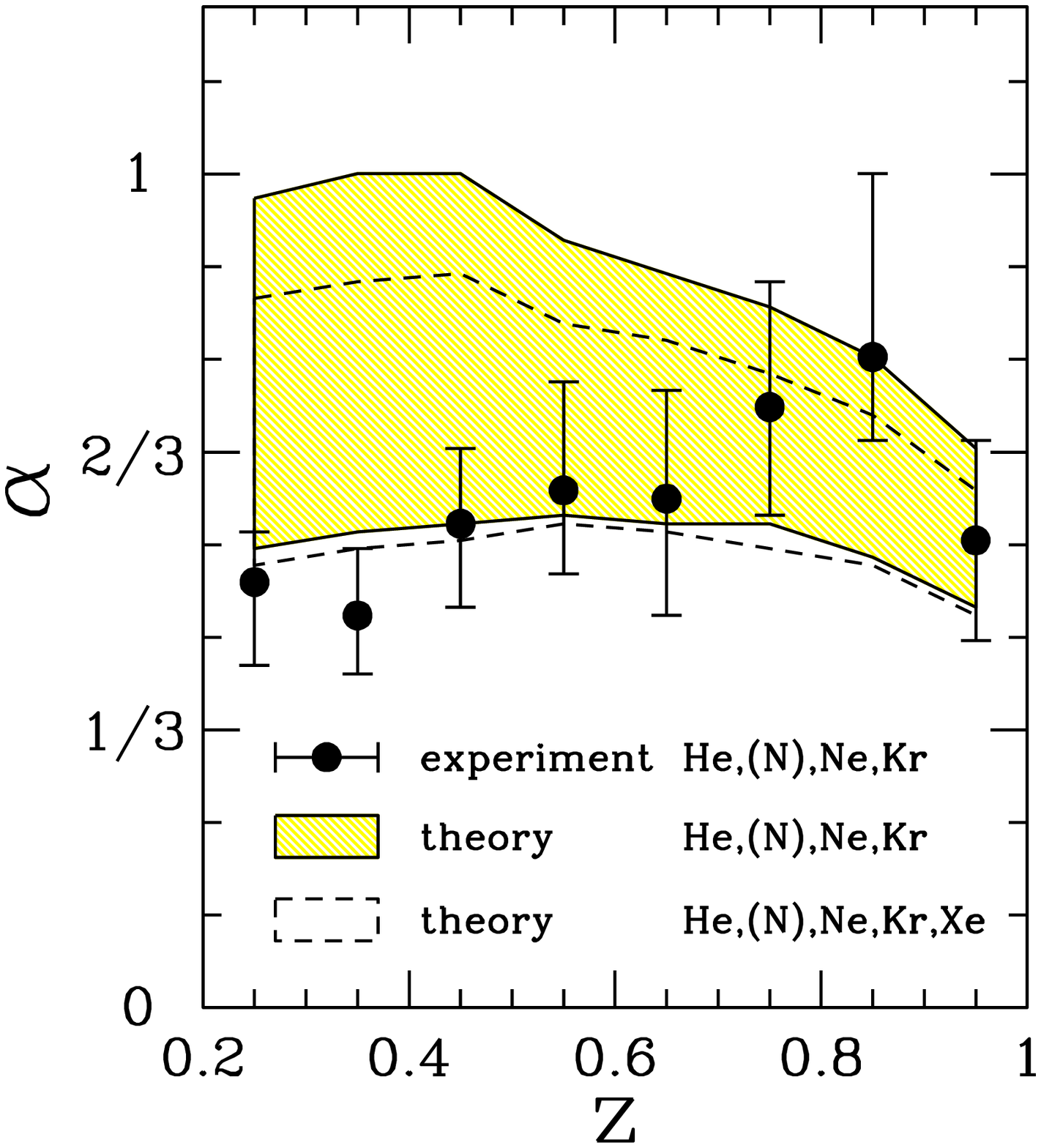}
\caption{Values of $\alpha$ as a function of $z$ derived from
  experimental data (dots) and our pure absorption model computation (bands).
  The nuclei included in the fit are shown  in the legend (the
  HERMES data on He and Ne are preliminary). The notation (N)
 indicates that this nucleus is included in the fit only
  at $z\geq 0.55$. Note that experimental errors are uncorrelated, but
  theory errors are point-to-point
  correlated. The experimental data and error bars as well as the theory band of $\alpha$ values for fits including the Xe nucleus have been corrected 
  compared to \cite{Accardi:2005jd}.
 }
\label{fig:fig9}
\end{figure}

A mistake in the computer program performing the power law fit of the
numerical computation of the hadron attenuation ratio $R_M$ has been
detected. The mistake affects all fits which include the Xe nucleus.
Below we present corrected results for table~\ref{table:fitcoeff} and
Figs.~\ref{fig:fig7}-\ref{fig:fig9}. 

Based on the corrected calculation we revise our conclusion in 
ref.~\cite{Accardi:2005jd}. The  $A^{2/3}$ power law
for $1-R_M$ in the absorption model remains also after including 
the Xe nucleus in the (c,$\alpha$) fit.  

\newpage
\begin{table}[t]
\begin{center}
\begin{tabular}{|c|c|c|} \hline
    &  \multicolumn{2}{c|}{Theory} \\
    &  \multicolumn{2}{c|}{He (N) Ne Kr Xe} 
    \\ \cline{2-3}
    $z$ & $c\;[10^{-2}]$ & $\alpha$
\\ \hline \hline
$.25$ &  $0.9\pme{0.9}{0.4}$  & $0.70\pme{0.15}{0.17}$ \\ \hline

$.35$ &  $0.8\pme{0.9}{0.4} $ & $0.72\pme{0.15}{0.17}$ \\ \hline

$.45$ &  $0.8\pme{0.9}{0.4}$  & $0.73\pme{0.15}{0.17}$ \\ \hline

$.55$ &  $0.9\pme{0.7}{0.4}$  & $0.71\pme{0.11}{0.13}$  \\ \hline

$.65$ &  $1.1\pme{0.8}{0.4}$  & $0.70\pme{0.10}{0.13}$ \\ \hline

$.75$ &  $1.4\pme{0.9}{0.4}$  & $0.68\pme{0.08}{0.13}$ \\ \hline

$.85$ &  $1.9\pme{1.2}{0.5}$  & $0.65\pme{0.06}{0.12}$ \\ \hline

$.95$ &  $3.3\pme{1.6}{0.7}$  & $0.57\pme{0.05}{0.10}$ \\ \hline
\end{tabular}
\vskip.3cm
\end{center}
\caption{Centroids of the contour plots  in
  Fig.~\ref{fig:fig7} with their uncertainties for
  the fit $1-R_M=c(\nu,z,h) A^\alpha$ at fixed $z$ bins.
  The correction affects only the fit with the Xe nucleus.}    
\label{table:fitcoeff}
\vskip.3cm
\end{table}

\bibliographystyle{unsrt}

\end{document}